\documentclass[aps,prl,twocolumn,showpacs]{revtex4}
\usepackage{graphicx}
\usepackage{color}

\begin{document}
\title{Angular dependence of the magnetic-field driven superconductor-insulator transition in thin films of amorphous indium-oxide}
\author{A.~Johansson}
\author{N.~Stander}
\author{E.~Peled}
\author{G.~Sambandamurthy} 
\author{D.~Shahar}
\affiliation{Department of Condensed Matter Physics, The Weizmann Institute of Science, Rehovot 76100, Israel}

\date{\today}

\begin{abstract}
A significant anisotropy of the magnetic-field driven superconductor-insulator transition is observed in thin films of amorphous indium-oxide. The anisotropy is largest for more disordered films which have a lower transition field. At higher magnetic field the anisotropy reduces and even changes sign beyond a sample specific and temperature independent magnetic field value. The data are consistent with the existence of more that one mechanism affecting transport at high magnetic fields.

\end{abstract}

\pacs{74.25.Fy, 74.78.-w, 74.25.Qt, 73.50.-h}

%74.25.Fy	Properties of type I and type II superconductors - Transport properties (electric and  	%			thermal conductivity, thermoelectric effects, etc.)
%74.78.-w	Superconducting films and low-dimensional structures
%74.25.Qt	Properties of type I and type II superconductors - Vortex lattices, flux pinning, flux creep
%73.50.-h		Electronic transport phenomena in thin films

\maketitle

The application of a strong magnetic field ($B$) perpendicular to the plane of a disordered superconducting film can drive it into an insulating state. In so-called uniformly disordered films this insulating state was argued \cite{Fisher1990} to result from a transition of the vortices induced by $B$ into a condensed state, forcing the Cooper-pairs to become localized. While several of the ingredients leading to this theory have received support in experiments \cite{Hebard1990,Yazdani1995b,Christiansen2002}, the existence, and role, of vortices in high-disorder thin-film superconductors near the SIT have not been verified in a direct observation. 

More troubling is the fact that a superconductor-insulator transition, driven by parallel $B$ \cite{Gantmakher2000b}, exhibits a surprisingly similar phenomenology to the more prevalent perpendicular $B$ transition. Studying amorphous Indium-Oxide (a:InO), Gantmakher et al. demonstrated that a transition $B$ ($B_{C}^{||}$) exists with the familiar temperature-independent resistance that sharply separates the superconducting and insulating behaviors. This $B_{C}^{||}$ was at a significantly higher value than $B_{C}^{\perp}$, the transition $B$ at perpendicular orientation. A scaling analysis of the data near the transition resulted in a critical exponent that is, within experimental error, equal to the one obtained at perpendicular $B$. The authors conclude that the two transitions belong to the same universality class, and argue that because in parallel orientation boson-vortex duality is not present, the theoretical approach of Ref. \cite{Fisher1990} is irrelevant to real systems.

The main purpose of this Letter is to present evidence, obtained from studying several a:InO thin-film superconductors, that there are (at least) two $B$-dependent mechanisms that affect electron transport near the SIT. The first mechanism dominates at lower $B$ and shows clear orientation dependence, and the second operates at higher $B$, and is closer to being isotropic. A $B$-driven SIT can take place in low-$B$ regime were transport is entirely dominated by the first, strongly anisotropic, mechanism and the contribution of the high-$B$ mechanism is negligible. 

To facilitate our study we grew a set of 30 nm-thick a:InO films by e-gun evaporation of 99.999\% sintered In$_{2}$O$_{3}$ targets onto glass substrate. Following a method described in details elsewhere \cite{Fiory1983,Sambandamurthy2004}, we heat treated our samples to achieve various degrees of disorder. The resulting films, measured at low temperatures ($T$s) and high $B$, have a critical temperature ($T_{c}$) in the interval 0.5 to 3 K, and exhibit clear $B$-driven transitions into an insulating state with $B_{c}$ varying between 0.125 and 11 T. The resistive response was studied in an Oxford dilution refrigerator attaining a base temperature of 10 mK, and equipped with a superconducting magnet capable of generating a $B$ of 12 T. The sample stage is equipped with a computer controlled Swedish Rotator, allowing us to control the angle between the surface of the film and the applied $B$ with an accuracy of 0.1 degrees.

We estimated the superconducting coherence length, $\xi$, using $\xi=\sqrt{\hbar D/k_{B}T_{c}}$, where $\hbar$ is Planck's constant, $k_{B}$ is the Boltzmann constant, and $D$ is the diffusion coefficient in the normal state calculated from the Einstein relation using a charge-carrier density of $5 \times 10^{20}$ cm$^{-3}$ \cite{Ovadyahu1981,Shahar1992}. The estimated $\xi$'s for our set of samples are found in the interval of 21-38 nm, indicating that our 30 nm films straddle the 2D limit for superconductors.
%Instead, it was estimated form the dirty limit expression \cite{TinkhamIntroSC}, $\xi=0.855\sqrt{\xi_{0}l}$, where $\xi_{0}$ is the Pippard coherence length and $l$ is the electronic mean free path. 

\begin{figure}[h]
\includegraphics[width=7.5cm]{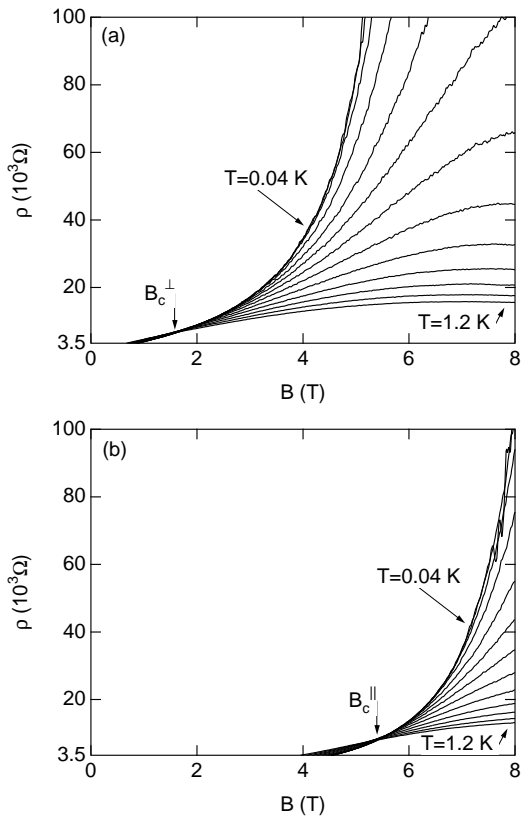}
\caption{Sheet resistance ($\rho$) as a function of $B$, from sample
Ba12/30, measured at $T=$ 0.04, 0.1, 0.2, 0.3, 0.4, 0.5, 0.6, 0.7, 0.8, 0.9, 1.0, 1.1, and 1.2~K. $B$ is
applied perpendicular (a) or parallel (b) to the surface of the film. The current in (b) is at approximately 45 degrees orientation to $B$. 
}
\label{SITAngle_R_B_T}
\end{figure}

A typical SIT transition, driven by a $B$ that is applied perpendicular to the film's surface, can be seen in Fig. \ref{SITAngle_R_B_T}(a) where we plot $\rho$ vs. $B$ for several $T$'s from 0.04-1.2 K. We identify the $B$ value where the different isotherms cross, and $\rho$ is independent of $T$, with $B_{c}^{\perp} = 1.59$ T, indicated by a vertical arrow. 

\begin{figure}
\includegraphics[width=7.5cm]{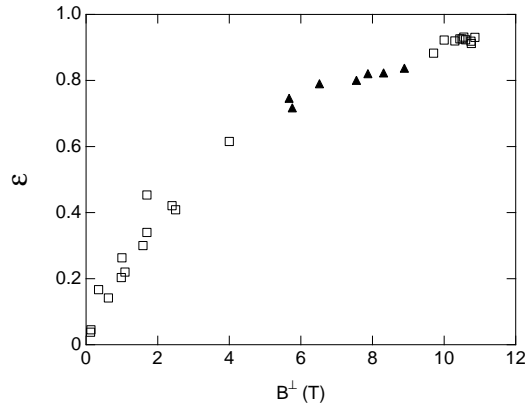}
\caption{Anisotropy factor $\epsilon$ as a function of corresponding
$B^{\perp}$ for our set of samples. Empty squares show
$B_{c}^{\perp}/B_{c}^{||}$ vs $B_{c}^{\perp}$, while filled triangles
show $B_{p}^{\perp}/B_{p}^{||}$ vs $B_{p}^{\perp}$.}
\label{SITAngle_epsilon}
\end{figure}

We then repeated our measurements, using the same film, this time rotating, {\em in-situ}, its plane such that the applied $B$ was parallel to the film's surface (Fig. \ref{SITAngle_R_B_T}(b)). Although not as well-defined, an SIT is observed with $B_{c}^{||} = 5.3$ T, a significantly higher value than $B_{c}^{\perp}$. We define the anisotropy factor, $\epsilon=B_{c}^{\perp}/B_{c}^{||}$ and, in Fig.  \ref{SITAngle_epsilon}, plot it vs. $B_{c}^{\perp}$ for our superconducting samples (empty squares). In light of the results of Gantmakher {\em et al.} \cite{Gantmakher2000b}, we were expecting a large $\epsilon$ in the transition $B$ for each of our samples.

While a significant shift was observed in films which had a low value of $B_{c}^{\perp}$, films with higher $B_{c}^{\perp}$ exhibited much smaller $\epsilon$. As can be seen in Fig. \ref{SITAngle_epsilon}, for films with higher $B_{c}^{\perp}$, which have a lower level of disorder, $\epsilon$ approaches unity, indicating a nearly isotropic $B_{c}$. 

\begin{figure}
\includegraphics[width=7.5cm]{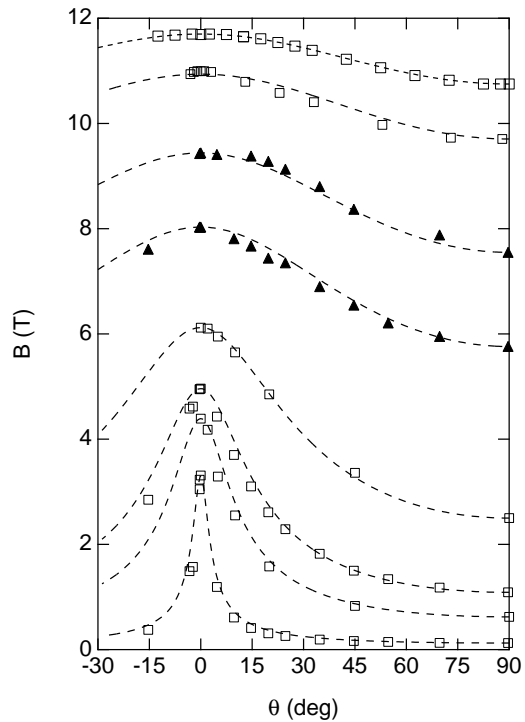}
\caption{Markers show measured $B_{c}$ (empty squares) or $B_{p}$ (filled triangles) vs. orientation of $B$ with respect to the plane of the film. Dashed lines correspond to fits to Eq. \ref{LDequation}.}
\label{SITAngle_Bc_rot}
\end{figure}

To further investigate the orientation dependence of $B_{c}$, we followed its evolution as a function of the angle of $B$ with respect to the plane of our films, $\theta$. The results, plotted in Fig. \ref{SITAngle_Bc_rot}, show that if one film has a $B_{c}^{\perp}$ larger than another, it will also have a larger $B_{c}^{||}$, although by a smaller factor. In other words, $B_{c}^{||}$ shows a weaker dependence on the level of disorder than $B_{c}^{\perp}$. 

The angular dependence of $B_{c}$, $B_{c}(\theta)$, for the SIT was considered by Meir and Aharony \cite{unpublMeirAharony}, using a percolation approach to the SIT. Within their model they predict the following form for $B_{c}(\theta)$:

\begin{equation}
	[\frac{B_{c}(\theta)\sin\theta}{B_{c}^{\perp}}]^2 +[\frac{B_{c}(\theta)\cos\theta}{B_{c}^{||}}]^2=1.
	\label{LDequation} 
\end{equation}

The dashed lines in Fig. \ref{SITAngle_Bc_rot} are plots of this function, without any fitting parameters except for the $B_{c}$ values obtained from the isotherms crossing-points. Incidentally, a similar form was derived by Lawrence and Doniach \cite{Lawrence1971} who modified, for layered superconductors, the earlier work of Tinkham and Harper \cite{Tinkham1963,Harper1968} that was based on solving a linearized Ginzburg-Landau equation applicable near $H_{c2}$. We emphasize that, for our samples, $B_{c}$ is substantially lower than $H_{c2}$ \cite{Sambandamurthy2004b}. 

\begin{figure}
\includegraphics[width=7.5cm]{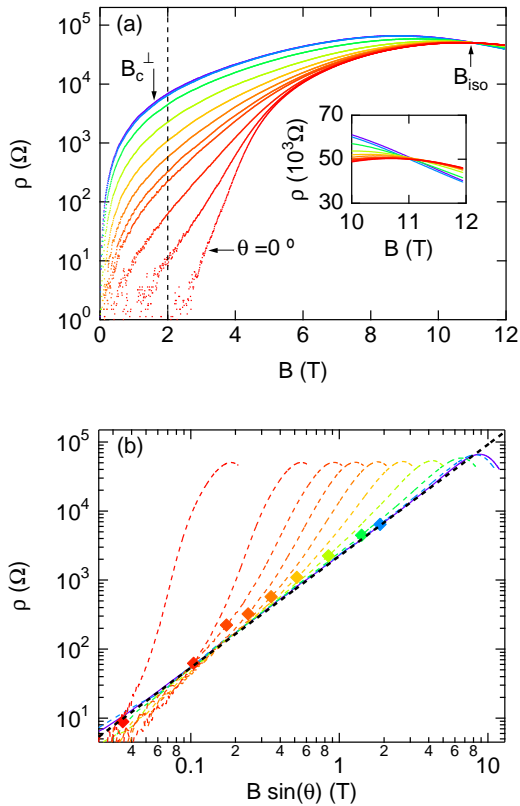}
\caption{(a) $\rho$ vs. $B$ from sample Ba12/30. The traces differ in angle $\theta$ between the plane of the sample and $B$. The $\theta$'s are 0 (red), 1, 3, 5, 7, 10, 15, 25, 45, 70, and 90 (purple) degrees. All traces were taken at $T$=0.6 K. Inset: $\rho$ is isotropic at the crossing point at $B$=11.02 T. (b) $\rho(\theta)$ replotted vs. $B \sin(\theta)$. The $\theta$'s are from the left 1 (red), 3, 5, 7, 10, 15, 25, 45, 70, and 90 (purple) degrees. $\rho(90)$ is fitted to a power law dependence (black, dashed line). Diamonds mark where total $B$ is 2 T, corresponding to the vertical dashed line in (a).}
\label{SITAngle_R_B_Angle}
\end{figure}

The diminishing anisotropy for samples with higher $B_{c}^{\perp}$ (and $T_{c}$) appears, on first sight, to be a result of the decreasing amount of disorder and the consequential strengthening of superconductivity. However, a detailed study of the angular dependence of $\rho$ in a given sample reveals a different picture. In Fig. \ref{SITAngle_R_B_Angle}(a) we plot the resistive response to $B$ applied at several values of $\theta$, obtained from one of our samples. The data presented are at a relatively high $T$ of 0.6 K avoiding, for the moment, the strong insulating peak that appear at low $T$ in disordered superconductors beyond the SIT \cite{Paalanen1992,Gantmakher1998,Butko2001,Sambandamurthy2004}. The data, spanning from perpendicular to parallel $B$, show a strong anisotropy at lower $B$ that is successively weakened as $B$ increases until, at a sample specific and well-defined $B_{iso}(=11.02$ T for this sample), $\rho$ is independent of $\theta$. At $B>B_{iso}$, the dependence on angle is reversed and $\rho$ has a larger value for parallel $B$ than for perpendicular $B$. The weakening (and reversal) of the anisotropy with increasing $B$ raises the possibility of reinterpreting the diminishing anisotropy in Fig. \ref{SITAngle_epsilon} as resulting, not from the decrease in disorder but, from the increase of $B$ where the transition takes place: Even in a given sample, with a fixed amount of disorder, the anisotropy diminishes at high $B$.

Although not sharply defined, we can identify two regimes with respect to the $\theta$ dependence of the magnetoresistance: A low-$B$ regime where $\rho$ is strongly anisotropic and a high-$B$ range where the anisotropy is not pronounced and even changes sign. Somewhat arbitrarily we use the $B$ at which we first detect resistance for $\theta=0$ as the border between the two regimes. We next consider these two regimes separately. 

Restricting the discussion to $B<2$ T, to the left of the vertical dashed line in Fig. \ref{SITAngle_R_B_Angle}(a), we find that $\rho(\theta)$ depends only on the perpendicular component of $B$. We demonstrate this in Fig. \ref{SITAngle_R_B_Angle}(b) by replotting $\rho (B)$ in (a) as a function of $B \sin(\theta)$. The trace taken in perpendicular $B$ is plotted as a purple, solid line and displays a good fit to an earlier reported \cite{Sambandamurthy2004} power law dependence on $B$ (black, dashed line). The remaining traces with a $B$ dependence rescaled by $\sin(\theta)$ are shown as dashed curves with the color code as in Fig. \ref{SITAngle_R_B_Angle}(a). The point at which the total $B$ equals 2 T is marked for each trace by a diamond with the same color as the trace itself. Plotted this way, the $\rho(B)$ traces taken at different $\theta$ values follow, within error, the $\rho (B)$ data for perpendicular field, demonstrating $\rho$'s dependence on the normal component of $B$ alone. 

Upon increasing $B$ for the sample in Fig. \ref{SITAngle_R_B_Angle}(b) beyond 2 T, the data no longer collapse on the perpendicular resistance curve. It is clear that one needs to introduce the effect of the non-vanishing parallel component. Trying to do this we encounter two difficulties. First, the parallel component can not be simply introduced in a fashion similar to the perpendicular component because it does not scale with $\cos(\theta)$ as we might expect if it was orbital in nature. In fact, the best fit to the data was obtained by assuming that the resistance that we measure at parallel $B$ (when the perpendicular contribution vanishes within error) is an isotropic component that contributes at all $B$ orientations. That brings us to the second difficulty: attempting a decomposition into an isotropic and anisotropic (proportional to $\sin(\theta)$) components yields only approximate success pointing to the possibility that the two mechanisms are not independent.

By dividing our $B$ range into two regimes we already alluded to the possibility that two distinct physical mechanisms are responsible for the resistance of our superconducting films in the presence of $B$.  At low $B$ a purely orbital component proportional to $\sin(\theta)$ is present, which have many characteristics that can be associated with vortices \cite{Sambandamurthy2004b}.  We stress that the SIT in perpendicular $B$ can take place in this low $B$ regime in which the parallel $B$ component is still negligible (see Fig. \ref{SITAngle_R_B_Angle}(a)). 

The higher $B$ range is dominated by the isotropic resistance component. In this range, at $B$ typically above 2-4 T in our samples, the system enters a strongly insulating peak at low $T$, that gives way to a lower resistance state at yet higher $B$ values. We next examine the $\theta$ dependence of the insulating peak.

\begin{figure}
\includegraphics[width=7.5cm]{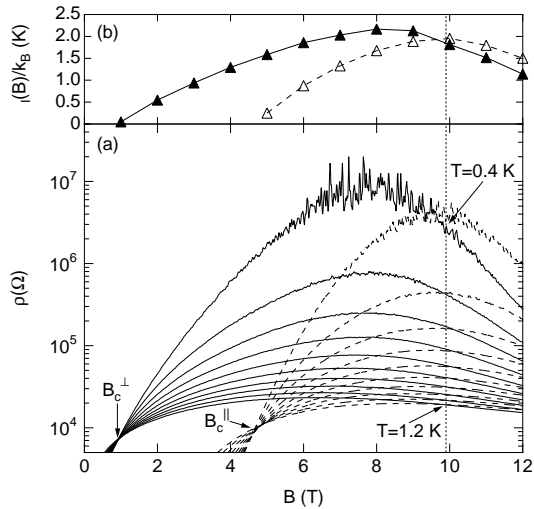}
\caption{(a) Isotherms of $\rho$ vs. $B$ from sample Ba12/70 for
perpendicular (solid lines) and parallel (dashed lines) $B$ orientation,
with $T$ ranging from 0.4 to 1.2 K in steps of 0.1 K. $B_{iso}$=9.9 T is
marked with a vertical dotted line. (b) Arrhenius fit to $R$ vs $T$ at
several $B$'s, displayed as filled triangles for perpendicular $B$ and
empty triangles for parallel $B$. Connecting lines are only guides for
the eye.}
\label{SITAngle_R_B_ParaPerp}
\end{figure}

In Fig. \ref{SITAngle_R_B_ParaPerp}(a) we plot isotherms of $\rho$ vs. $B$ for perpendicular and parallel (dashed lines) $B$. Both configurations exhibit the insulating peak, but its maximum is smaller for the parallel configuration. Considering the activated nature of transport in the insulating regime \cite{Sambandamurthy2004} we can extract $\Delta_{I}(B)$, the activation energy, for each configuration. Plotted in Fig. \ref{SITAngle_R_B_ParaPerp}(b) we see that the $\Delta_{I}(B)$'s have a rather similar behavior and magnitude, with the $\Delta_{I}(B)$ in the parallel configuration shifted with respect to the perpendicular one. This shift can actually be seen in the resistance data themselves in Fig. \ref{SITAngle_R_B_ParaPerp}(a). The similarity between the parallel and perpendicular configurations in this high $B$ insulating regime indicates that a mechanism that is largely insensitive to $B$ orientation is responsible for the insulating peak. The nature of this insulating peak is still a mystery \cite{Dubi2005}. 

When following the evolution of the position of the insulating peak, $B_{p}(\theta)$, as a function of $\theta$ (filled triangles in Fig. \ref{SITAngle_Bc_rot}), it shows a dependence close to that of $B_{c}(\theta)$ described by Eq. \ref{LDequation}. Interestingly, the anisotropy factor for the peak position, $B_{p}^{\perp}/B_{p}^{||}$, also shows ( Fig. \ref{SITAngle_epsilon}, filled triangles) a behavior compatible with the $B_{c}$ anisotropy, $\epsilon$.
 
Finally, inspecting Fig. \ref{SITAngle_R_B_ParaPerp}(a) reveals that $B_{iso}$ is nearly independent on $T$. This is true even when the value of $\rho$ itself at $B_{iso}$ is strongly (exponentially) $T$ dependent. While the significance of this $B$ is unclear, we note that for $B>B_{iso}$, $\rho(B)$ in the parallel orientation is larger than 
$\rho(B)$ when $B$ is applied normal to the film's surface.
 
%To summarize, we have shown a strong anisotropy present at low $B$ for disordered thin films of a:InO. The anisotropy is gradually weakened with increasing $B$, and eventually reversing its direction at a sample specific value. 
%To summarize, we have shown an anisotropy in $B_{c}$ for disordered thin films of a:InO. The anisotropy is strong for films with low $B_{c}$ and is gradually weakened with increasing $B_{c}$. From studying $\rho(\theta)$ at low and high $B$, we conclude that there is more than one resistive mechanism involved. We found that the strength of the anisotropy rather correlates to the strength of $B$ than the level of disorder in the films. The developing picture from the dissipative response supports aspects of  both presence of vortices and percolation in the films.

To summarize, we have shown an anisotropy in $B_{c}$ for disordered thin films of a:InO. The anisotropy is strong for films with low $B_{c}$ and is gradually weakened with increasing $B_{c}$. We found that the strength of the anisotropy rather correlates to the strength of $B$ than the level of disorder in the films. From studying $\rho(\theta)$ at low and high $B$, we conclude that there is more than one resistive mechanism involved. The developing picture from the dissipative response supports aspects of  both presence of vortices and percolation in the films. Finally, the data reveal a sample specific $B$ value, for which $\rho$ is isotropic and beyond which $\rho$ is larger in parallel $B$. 
 
%To conclude, we have shown an anisotropy in $B_{c}$ for disordered thin films of a:InO. The anisotropy is strong for films with low $B_{c}$ and is gradually weakened with increasing $B_{c}$. We found support for more than one resistive mechanism by considering the angular dependence of $\rho$ vs $B$, which is close to independent of angle at high $B$, while developing a strong anisotropy at lower $B$. Finally, the data show a sample specific $B$ value, for which $\rho$ is isotropic and beyond which $\rho$ is larger in parallel $B$. 

We wish to thank E. Altmann, Y. Meir, Y. Oreg, and Z. Ovadyahu for useful discussions. This work was supported by the ISF, the Koshland Fund and the Minerva Foundation. 

%\bibliography{biblio3}

\end{document}